# Oxy-acetylene driven laboratory scale shock tubes for studying blast wave effects.

Amy C. Courtney, Ph.D.[1,a], Lubov P. Andrusiv, Ph.D.[2], and Michael W. Courtney, Ph.D.[2]

[1] R & D, Force Protection Industries, Inc., 9801 Highway 78, Ladson, SC, 29456
[2] U.S. Air Force Academy, 2354 Fairchild Drive, USAF Academy, CO, 80840

This paper describes the development and characterization of modular, oxy-acetylene driven laboratory scale shock tubes. Such tools are needed to produce realistic blast waves in a laboratory setting. The pressure-time profiles measured at 1 MHz using high speed piezoelectric pressure sensors have relevant durations and show a true shock front and exponential decay characteristic of free-field blast waves. Descriptions are included for shock tube diameters of 27 - 79 mm. A range of peak pressures from 204 kPa to 1187 kPa (with 0.5 – 5.6% standard error of the mean) were produced by selection of the driver section diameter and distance from the shock tube opening. The peak pressures varied predictably with distance from the shock tube opening while maintaining both a true blast wave profile and relevant pulse duration for distances up to about one diameter from the shock tube opening. This shock tube design provides a more realistic blast profile than current compression-driven shock tubes, and it does not have a large jet effect. In addition, operation does not require specialized personnel or facilities like most blast-driven shock tubes, which reduces operating costs and effort and permits greater throughput and accessibility. It is expected to be useful in assessing the response of various sensors to shock wave loading; assessing the reflection, transmission, and absorption properties of candidate armor materials; assessing material properties at high rates of loading; assessing the response of biological materials to shock wave exposure; and providing a means to validate numerical models of the interaction of shock waves with structures. All of these activities have been difficult to pursue in a laboratory setting due in part to lack of appropriate means to produce a realistic blast loading profile.

## I. INTRODUCTION

The increased prevalence in recent years of blast-induced damage to materiel and injury to personnel has motivated laboratory scale experiments on effects of blast waves in hopes of improving armor design and mitigating or treating resulting damage or injuries.[1-7] Shock tubes have been used for over a century to study combustion chemistry, high speed aerodynamics, and shock wave characteristics as well as the response of materiel to blast loading. The earliest reports in English appeared at the turn of the 20$^{th}$ century; interest in research with shock tubes revived in the U.S. and Canada after World War II, and it was at that time that the term "shock tube" became prevalent.[8]

Current shock tubes include compression-driven and blast-driven designs of varying dimensions, with shock tube openings varying in diameter from a few centimeters to a meter or more and widely ranging peak pressures from 30 kPa to more than 60 MPa. These have been used to apply pressure waves to test subjects and materiel from small animal models to large pieces of equipment and vehicles. However, these compression-driven and blast-driven shock tubes suffer from drawbacks that limit their usefulness.

Compression-driven shock tubes exhibit significant shot to shot variations in peak pressure,[7] produce pressure waves with durations longer than typically encountered from real threats such as antipersonnel mines, hand grenades, and improvised explosive devices,[6,9,10] and do not accurately reproduce the Friedlander waveform of free-field blast waves.[3,4] Moreover, the expansion of the compressed gases results in a "jet effect" much larger than that produced by a blast wave. This jet of expanding gases applies additional force and transfers momentum to a test object. This secondary loading may be especially undesirable for studies of the primary effects of blast waves on biological specimens.

Blast-driven shock tubes produce more realistic pressure-time profiles, but their operation requires facilities, liability, and personnel overhead for storing and using high explosive materials. In addition, equipment and personnel need to be isolated from the large mechanical and electromagnetic waves caused by detonation.[1,2]

---

[a] Author to whom correspondence should be addressed. Electronic mail: amy_courtney@post.harvard.edu

In recent years, the need for laboratory scale tools to apply realistic blast waves to animal models has become apparent.[11] In addition to the challenge of producing realistic blast waves, there is also the challenge of isolating exposure to specific body regions. This is desirable to investigate plausibility of and thresholds for proposed mechanisms of blast-induced injury and to study wave propagation. Some are also interested in investigating the effects of shock waves on isolated tissues or cell cultures.[12] Most current shock tube designs do not meet all of the requirements for fidelity of the pressure wave and accessibility in terms of facilities and cost.[13]

The utility of a modular shock tube design has been demonstrated in both compression-driven and blast-driven designs. For example, in the 1950s Henshall[8] built a modular, compression-driven shock tube that consisted of mahogany wood channels with a square cross section. A decade later, Duff and Blackwell[14] described a modular, blast-driven shock tube constructed from low-cost, easily replaceable parts. Their designs ranged in diameter from 0.6 to 2 m and in length from 3 to 15 m and produced peak pressures between 7 MPa and 200 MPa.

Since the mid 1960's, shock tubes have regularly been used to investigate the chemistry of combustion of many fuel-air mixtures with the purpose of optimizing fuel and/or engine performance.[15]

In this study, a modular, oxy-acetylene driven, laboratory scale shock tube design was developed to meet the need for a laboratory scale shock tube that can apply realistic loading profiles to relatively small areas and without the high cost and effort required to construct and operate a shock tube using previously available designs. The design uses readily available components in a modular system. Shock tube diameters from 27 mm up to 79 mm were used to produce peak pressures in the range of 204 to 1187 kPa.

## II. MATERIALS AND METHODS

### A. Shock Tube Construction

Both driven and driver sections consist of commercially available steel pipe, and they are coupled by a steel flange. The driver section was sealed with a steel cap, into which a hole was drilled for ignition access (Figure 1). The internal seam due to the rolled construction of the pipe had no negative effects on the blast wave profile. The dimensions of several driver and driven components tested are listed in Table I.

A hole (labeled "sensor mount" in Figure 1) was drilled and tapped near the end of each driven section for mounting of a piezoelectric, high-speed pressure sensor (PCB 102B18) to measure pressure parallel to the direction of travel of the shock wave and near the shock tube opening.

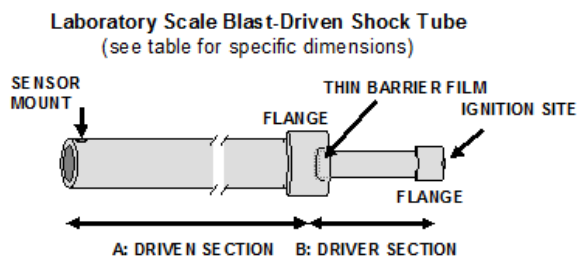

Figure 1: Components of the laboratory scale oxy-acetylene driven shock tube.

The length chosen for the driven sections was informed by development of a table-top shock tube driven by commercially available rifle primers.[16] Those results suggested that choosing the length of the driven section to be about 60 diameters resulted in a desirable blast loading profile. As it happens, the length of most of the driven sections tested were about ten times the length of the driver sections, which is consistent with early published shock tube designs.[8]

### B. Shock Wave Production

The driver section was filled with a stoichiometric mixture of oxygen and acetylene. Combustion products of this mixture are carbon dioxide and water vapor.

Prior to filling the driver section with the fuel-oxygen mixture, a thin barrier material was placed over the open end to contain the mixture, with a small ventilation tube placed parallel to the driver section to allow ambient air to escape during filling. A single layer of food-grade plastic film (low density polyethylene) held in place with a rubber band was a convenient and effective barrier. Two layers of teflon tape were applied to the threads of the driver section before and after placement of the plastic film barrier to prevent the threads from cutting the film prior to ignition.

TABLE I. Laboratory scale blast-driven shock tube component dimensions. The lettered designations correspond to the labeled regions in Figure 1. Sections are coupled by a steel flange.

| **A:** DRIVEN SECTION | 27 mm | 41 mm | 51 mm | 79 mm | | |
|---|---|---|---|---|---|---|
| Length (cm) | 183 | 305 | 305 | 305 | | |
| Inner diameter (cm) | 2.7 | 4.1 | 5.1 | 7.9 | | |
| Outer diameter (cm) | 3.4 | 4.9 | 5.9 | 8.8 | | |
| Mounted sensor center distance from opening (cm) | 1.1 | 1.2 | 1.2 | 1.0 | | |
| **B:** DRIVER SECTION | 16 mm | 21 mm | 27 mm | 41 mm | 51 mm | 79 mm |
| Length (cm) | 26.7 | 25.4 | 30.5 | 30.5 | 30.5 | 30.5 |
| Inner diameter (cm) | 1.6 | 2.1 | 2.7 | 4.1 | 5.1 | 7.9 |
| Outer diameter (cm) | 2.2 | 2.7 | 3.4 | 4.9 | 6.0 | 8.8 |

Latex balloons (classified as 12-inch balloons when inflated) secured by their own elasticity were also tested as a barrier material. For both barrier materials, the pressure-time profile of the shock wave was nearly identical, but for some trials when the balloon was used, a small piece of latex was propelled onto the target or sensor at speeds up to 200 m/s (determined using high-speed video analysis), producing a detectable impulse that may be undesirable for certain applications. In contrast, use of the plastic film resulted in much smaller particles (mass < 1 mg) that did not detectably interfere with pressure wave measurements.

Immediately after the driver section was filled via the ignition access, the ignition access was temporarily sealed with putty and the ventilation tube was removed. The ignition source, an electric match, was installed via the ignition access, which was then resealed with putty. The prepared driver section was threaded into the coupling flange and the leads to the ignition source were attached to a remote 9 V DC source.

*C. Characterization*

Blast wave characteristics were measured for combinations of driver and driven sections at several distances from the shock tube opening. In addition to the piezoelectric pressure sensor mounted at the end of the driven section, a second sensor (PCB 102B15) was placed facing the opening of the shock tube, with its central axis aligned with that of the shock tube. Five trials were recorded for each combination of driver/driven section tested and at each distance.

Pressure data were recorded at a sample rate of 1MHz via cables connecting each pressure transducer to a signal conditioning unit (PCB 842C), which produced a calibrated voltage output. The voltage output was then digitized with a National Instruments PXI-5105 or USB-5132 fast analog to digital converter. The voltage waveform was converted to pressure using the calibration provided by the manufacturer with each pressure sensor. Tests were conducted at about 15°C and an ambient air pressure of about 78 kPa.

### III. RESULTS

*A. Experimental Results*

The shock waves generated by each combination of driver and driven section tested had rise times to peak pressure of a few microseconds, an exponential decay, and pulse durations typical of free-field blast waves. Table II shows peak pressures measured for different configurations and distances from the shock tube opening. Results are shown for pressures measured face-on and parallel to the direction of travel of the shock wave. Five trials each of several driver/driven section combinations at different distances from the shock tube opening showed consistent results. Standard error of the mean peak pressure measured face-on was less than six percent for any of the conditions tested, and less than three percent for most combinations. Mean pressures measured parallel to the direction of travel include all trials for a given combination of driver and driven section, since the position of that sensor did not change. Standard error of the mean peak pressure measured parallel to the direction of travel was two percent or less.

The pressure-time plots in Figure 2 show typical results for the pressure-time profiles and variation in peak pressure obtained by use of two different driver sections on the same driven section. Figure 2 also illustrates that similar peak pressure can be achieved by more than one combination of driver and driven section. Peak pressure produced by a 51 mm driver in the 51 mm driven section were similar to those produced by the 79 mm driver in the 79 mm driven section. Of course, the total energy and area over which the pressure is applied is higher for the 79 mm driver in the 79 mm driven section.

In Figure 3 pressures measured parallel to the direction of propagation of the shock wave and face-on at the opening are compared. The peak pressure measured face-on is about three times greater than that measured parallel to the direction of travel. This is due to pressure reflected from the sensor face and the factor of about three is an expected magnitude for atmospheric conditions typically found across North America and Europe. The peak pressure measured by the sensor mounted at the end of the shock tube and parallel to the direction of travel of the shock wave was linearly correlated to the peak pressure measured by the sensor facing the shock tube ($R^2 > 0.99$; functions pass through the origin).

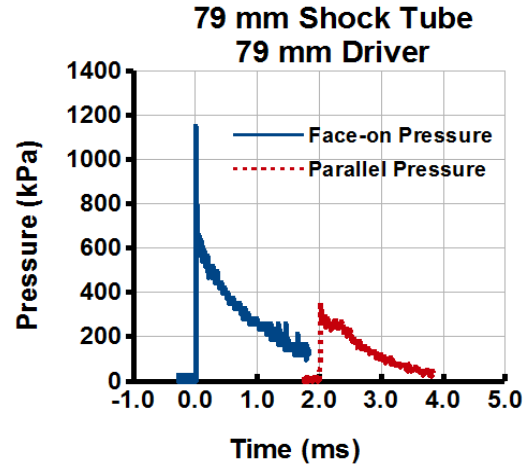

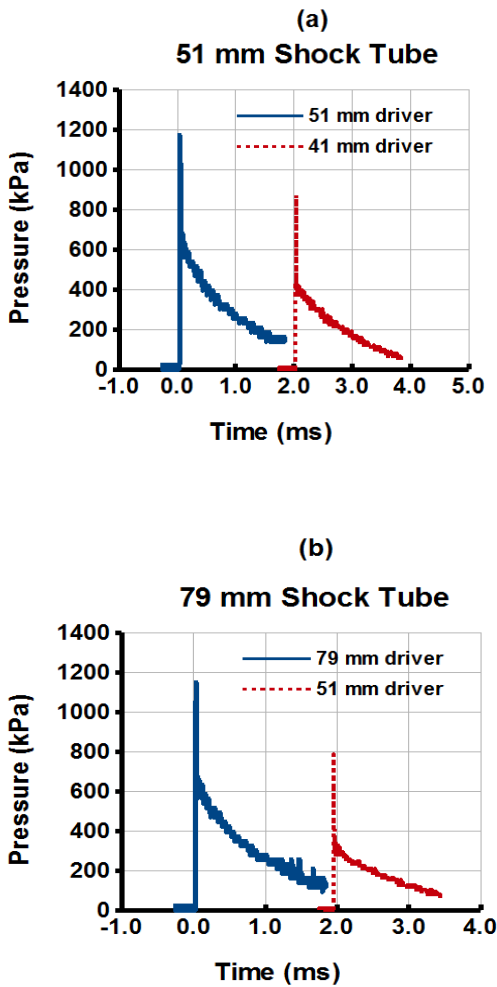

*Figure 2. Typical pressure vs. time waveforms produced by: (a) 51 mm driven section and 51 mm and 41 mm driver sections, (b) 79 mm driven section and 79 mm and 51 mm driver sections. Pressure shown was measured with the sensor facing the shock tube at the opening. Waveforms are offset in time for ease of viewing.*

*Figure 3. Pressure vs. time waveforms measured face-on and parallel to the direction of travel of the shock wave produced by a 79 mm driven section with a 79 mm driver section. Waveforms are offset in time for ease of viewing.*

Only when the diameter of a driver section was smaller than that of the driven section, a secondary, smaller peak pressure was measured 1.5 – 2.0 ms after ignition occurred, depending on the specific configuration (Figure 4). This secondary peak was detected by both sensors. However, it did not occur when the diameters of the driven and driver sections were the same. Based on these observations, the length of the driver section, and the speed of propagation of the shock wave, this small secondary rise in pressure is probably due to constructive interference of pressure waves reflecting at the coupling, then off the back of the driver section, and finally continuing down the driven section.

Experiments were performed at an ambient air pressure of about 78 kPa due to high altitude. The oxy-acetylene shock tubes are operated at ambient pressure. Therefore, at higher air pressures, a greater number of moles of the oxy-acetylene mixture will be contained by a given driver section, and a higher peak pressure can be expected. Tests performed at sea level with the 27 mm and 41 mm driven sections support this expectation (unpublished data). The difference in peak pressure is expected to be approximately proportional to the difference in ambient air pressure, but each system should be characterized individually.

TABLE II. Peak pressures generated by the oxy-acetylene driven laboratory scale shock tube measured for different configurations and distances from the shock tube opening. Peak face-on pressures at each distance are the mean of five trials; peak parallel pressures for each configuration are the mean of 15-20 trials. Uncertainty is reported as standard error of the mean (SEM).

| Driven Section Diameter (mm) | Driver Section Diameter (mm) | Distance from Opening (mm) | Peak Face-on Pressure (kPa) (n=5) | SEM (kPa) | Peak Parallel Pressure (kPa) (n=15-20) | SEM (kPa) |
|---|---|---|---|---|---|---|
| **27** | 16 | 0 | 611 | 25 | 217 | 4 |
| | | 20 | 480 | 10 | | |
| | | 40 | 204 | 6 | | |
| | 21 | 0 | 920 | 13 | n/a | n/a |
| | | 20 | 824 | 6 | | |
| | | 40 | 296 | 6 | | |
| **41** | 21 | 0 | 347 | 5 | n/a | n/a |
| | 27 | 0 | 588 | 13 | 180 | 2 |
| | | 20 | 623 | 35 | | |
| | | 40 | 405 | 25 | | |
| **51** | 41 | 0 | 900 | 15 | 297 | 3 |
| | | 20 | 918 | 31 | | |
| | | 40 | 812 | 15 | | |
| | | 60 | 521 | 13 | | |
| | 51 | 0 | 1163 | 14 | 343 | 4 |
| | | 20 | 1149 | 17 | | |
| | | 40 | 982 | 24 | | |
| | | 60 | 679 | 16 | | |
| **79** | 51 | 0 | 790 | 5 | 236 | 2 |
| | | 20 | 776 | 12 | | |
| | | 40 | 771 | 4 | | |
| | | 60 | 736 | 14 | | |
| | 79 | 0 | 1163 | 11 | 351 | 5 |
| | | 20 | 1187 | 25 | | |
| | | 40 | 1173 | 22 | | |
| | | 60 | 1069 | 23 | | |

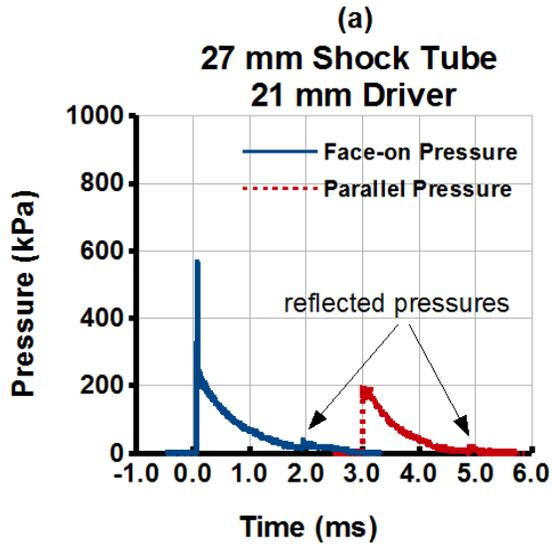

Figure 4. Pressure vs. time waveforms measured face-on and parallel to the direction of travel of the shock wave produced by: (a) 27 mm driven section with a 21 mm driver, and (b) 79 mm driven section with a 51 mm diameter driver section. The small secondary peaks are due to internal reflection. This occurred only when the diameter of the driver section was smaller than that of the driven section. Waveforms are offset in time for ease of viewing.

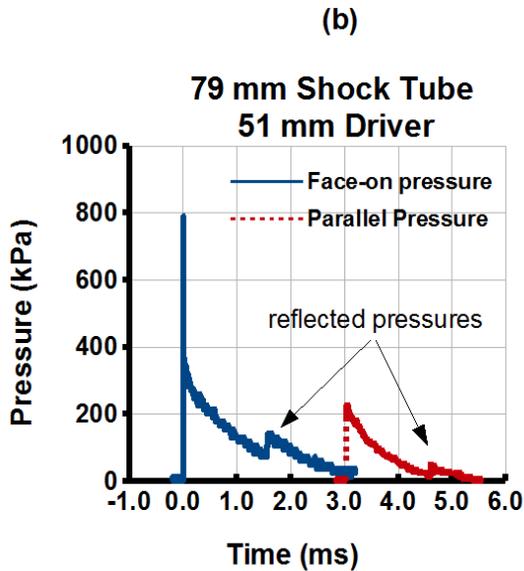

*B. Interpolation*

The results can be used to determine the peak pressure that would be applied to a test sample by placing it at intermediate distances from the shock tube opening. The interpolation functions are of the following form:

$$P(x) = \frac{a}{\left(\frac{x}{x_0}\right)^b + 1}, \quad (1)$$

where $P(x)$ is the average peak pressure measured in kPa at distance $x$ mm from the shock tube opening. Table III lists the computed parameters. The parameters have physically relevant meanings – $a$ is approximately the peak pressure at the tube opening, $b$ is the power describing the rate of decrease in peak pressure with distance from the opening, and $x_0$ is the distance at which the peak pressure is half that at the opening.

*TABLE III. Parameter values for interpolation functions describing falloff of peak pressure with distance from the opening for several combinations of driver and driven sections of the laboratory scale shock tube. For each relationship, $R^2 > 0.99$.*

| Driven Section Diameter (mm) | Driver Section Diameter (mm) | $a$ | $b$ | $x_0$ |
|---|---|---|---|---|
| 27 | 16 | 611.0 | 2.87 | 31.00 |
|    | 21 | 920.3 | 4.18 | 33.44 |
| 41 | 27 | 604.3 | 4.00 | 47.88 |
| 51 | 41 | 910.6 | 4.54 | 63.94 |
|    | 51 | 1166.0 | 3.34 | 66.20 |
| 79 | 51 | 785.4 | 2.66 | 166.60 |
|    | 79 | 1184.2 | 3.00 | 129.49 |

Interpolation functions for peak pressure as a function of distance from the shock tube opening are plotted along with experimental data in Figure 5. The plots illustrate that the peak pressure is maintained for longer distances from the shock tube opening for larger diameter driven sections. Also, the same peak pressure can be produced by more than one combination of driver and driven sections. The total energy contained in the shock wave and the area over which the shock wave is applied varies with the diameters of the driver and driven sections, of course, and a certain combination may be more suitable for a given application.

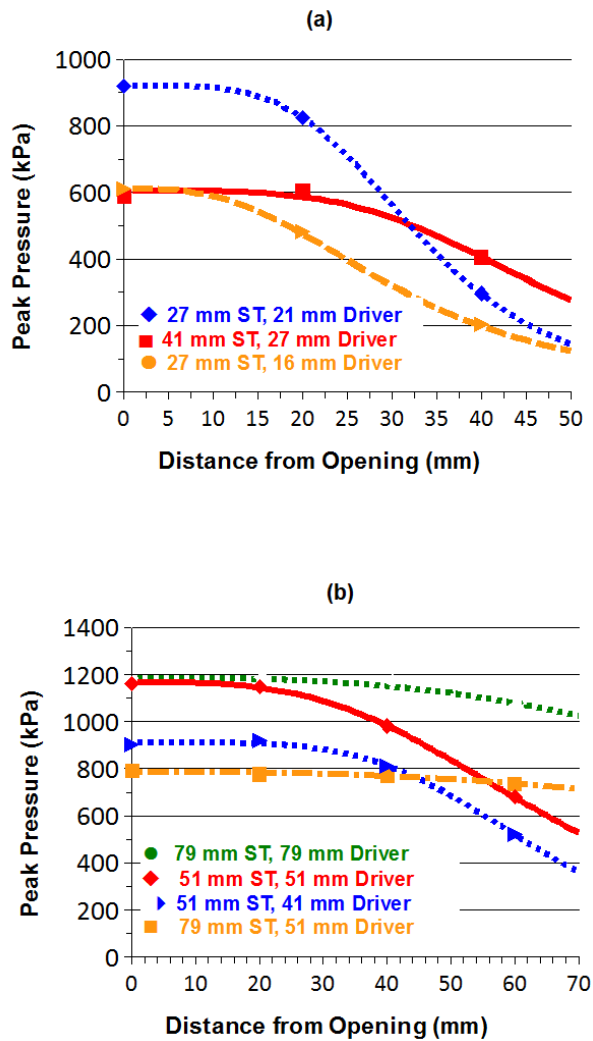

*Figure 5. Interpolation functions (lines) of the form of Equation (1) were fit to the data (solid symbols) to facilitate selection of specimen placement for exposure to intermediate peak pressures. Peak pressure as a function of distance from the shock tube opening is shown for some configurations of the laboratory scale shock tube using a) 27 mm or 41 mm driven sections, b) 51 mm or 79 mm driven sections. Note the faster falloff of peak pressure with distance for smaller diameter driven sections.*

## IV. DISCUSSION

In a compression driven shock tube, the peak pressure applied to a test object is determined by the pressure in the driver section prior to release. The pressure in the driver section must be measured to reduce shot-to-shot variation. In contrast, the laboratory scale, oxy-acetylene driven shock tubes are operated at ambient pressure. The peak pressure applied to a test object is repeatable and can be adjusted by the choice of driver and driven sections as well as the distance between the shock tube opening and the test object. Only some of the possible combinations of this modular design have been characterized in the present study. The goal was to describe the approach in enough breadth and depth to equip investigators to identify useful combinations for specific applications. With similar components, similar results can be expected, with adjustment for any difference in ambient pressure. While the results can be used to guide component selection, each system should be characterized before being used for experiments.

The modular design has proven to be robust in that several combinations of diameters of driver and driven sections consistently produced pressure-time profiles like those of free-field blast waves. It may be that larger diameter driven sections than described herein will require a longer length to maintain a desirable pressure-time profile. Because of the strong correlation between peak pressures measured by the mounted and external pressure sensors, the sensor mounted at the end of the shock tube can be used to verify the pressure applied to a test object for each exposure once a configuration has been characterized.

These laboratory scale shock tubes were developed with several applications in mind. They are expected to be useful for applying a blast wave to candidate armor materials to determine transmission, reflection and absorption properties. They can be used to apply a blast wave to biological samples and to localize exposure to a region of a larger specimen. They can also be used to determine material properties at blast strain rates.

On a more basic level, these shock tubes can be used to assess and compare responses of pressure, force and strain sensors of different types and in different configurations prior to application in high rate experiments where the loading is not well characterized or may not be as repeatable.

Experiments like those suggested above can also be used with simple structures to verify numerical modeling techniques, in order to increase confidence in the quantitative results produced by those techniques when they are applied to geometrically and mechanically complex structures. Such experimental validation of numerical models is presently an important and pressing need in the understanding of blast effects.

By selection of components and distance from the shock tube opening, effects of increasing the exposure area for the same peak pressure can be studied. At the distances tested, several configurations produced pulse durations of 2 ms; therefore this design can also be used to investigate the effects of increased peak pressure at comparable pulse durations. Published data from current designs show that the pulse duration increases with increased peak pressure, making it difficult to determine how each quantity affects the response of the object under test.

Reneer et al.[5] described a multi-mode shock tube driven either by compressed air or by oxy-hydrogen and cyclotrimethylenetrinitramine (RDX). Consistent with results from other shock tubes, they reported that shock waves produced by compressed air had durations exceeding those produced by RDX. The design can thus be used to study the effect of increasing positive pulse duration while keeping the peak pressure constant. However, like other blast-driven designs, it requires significant investment and specialized facilities and personnel due to the use of an explosive and to some undesirable combustion products.

Although the combustion temperature of oxy-acetylene is high, thermal damage to a test object from the oxy-acetylene driven shock tube is not expected. The oxy-acetylene mixture is contained at ambient pressure in the 30 cm long driver section prior to ignition. The molar volume of the combustion products is smaller than the molar volume of the reactants. Therefore, there is neither a large expansion of the fuel gas nor a flame emerging from the 183 cm or 310 cm long (depending on the diameter) driven section of the shock tube after ignition. The expanding, gaseous products of the combustion are cooling as they travel down the driven section of the shock tube, and the interaction time between the expanding gases and the test object is short, so that heat transfer is minimal. In tests where paper, plastic, leather and bone have been placed 1-3 cm from the shock tube opening, no evidence of thermal damage (such as melting, singeing or discoloration) has been observed even after several exposures. No part of the shock tube is hot and any part of the shock tube can be easily handled right after a test.

A limitation of compression driven shock tubes is the "jet effect" that follows the shock wave and which is produced by the expansion of the compressed gases. This jet of expanding gases applies additional force and transfers momentum to the test object. This additional loading may be undesirable, especially when the effects of blast waves on biological specimens are being studied. The oxy-acetylene driven shock tube does not produce a large jet effect. To support this assertion, calculations are presented in the Appendix estimating the jet effect produced by oxy-acetylene and by the amount of RDX and compressed gas required to produce the same energy in the same driver section.

**V. CONCLUSIONS**

In response to the need for tools that can produce realistic blast waves in a laboratory setting[12, 14, 17] and be useful for testing the response of materiel, biological materials and sensors to blast waves, a modular laboratory scale shock tube was designed. Unlike current shock tube designs, this design can be used to apply true shock waves with realistic profiles to small areas of a test subject or candidate armor material. By varying the dimensions of easily interchangeable driver and driven sections, shock tubes of diameters 27 mm to 79 mm were used to produce peak pressures ranging from about 204 to 1187 kPa. The peak pressures varied predictably with distance from the shock tube opening while maintaining a true blast wave profile and relevant positive pulse duration for distances about one diameter from the shock tube opening.

These shock tubes can be built with standard steel pipe stock and fittings and operated by non-EOD personnel in less specialized facilities than required by current blast-driven shock tubes, greatly reducing overhead cost and effort and increasing throughput and accessibility. Several types of experimental applications have been described.

**ACKNOWLEDGMENTS**

This work was supported in part by BTG Research, www.btgresearch.org.

**APPENDIX**

Calculations were performed to compare the amount of additional gas produced by the oxy-acetylene, RDX and compressed air required to produce 1292 J, the same energy as produced in experiments with the oxy-acetylene laboratory scale shock tube using a 5.1 cm x 30.5 cm driver section (V = $6.2306 \times 10^{-4}$ m$^3$) and performed at an air pressure of 78,008.7 Pa (measured with a Kestrel digital weather meter) and temperature of 293°K (68°F). Under these conditions, the molar volume of an ideal gas is 31.2291 liters. Results are summarized in Table AI.

Calculations were performed using the molecular weights and heats of formation for reactants and products of the respective combustion or explosion equation, along with the ideal gas law. The combustion reaction for acetylene is

$$2C_2H_2 + 5O_2 \rightarrow 4CO_2 + 2H_2O \qquad (A1)$$

After combustion, the gas products occupy less volume by 89.01 cm$^3$ (-0.002850 moles). This result suggests that the oxy-acetylene shock tube does not have a large jet effect.

The explosion equation for RDX is

$$2C_3H_6N_6O_6 + 3O_2 \rightarrow 6H_2O + 6CO_2 + 6N_2. \qquad (A2)$$

First, the heat of explosion per mole of RDX was computed, then the number of moles required to produce 1292 J of energy was determined. RDX has a negative oxygen balance and will not react completely unless there is atmospheric oxygen available. Therefore, the number of moles of oxygen required was subtracted from the final number of moles of gas to compute the net increase.

The energy released when a compressed gas is released to atmospheric pressure is its pressure times its volume. The number of moles of compressed gas required to store 1292 J of energy in the specified volume was determined. The net number of moles of gas released was determined by subtracting the number of moles of gas in the driver section at atmospheric pressure.

TABLE AI. *Comparison of the number of moles of additional gas produced by different sources of a blast or shock wave in a laboratory scale shock tube (driver section 5.1 x 30.5 cm).*

| Source of blast or shock wave | Estimated moles of additional gas produced | % compared to moles of compressed gas required |
|---|---|---|
| Oxy-acetylene | -0.002850 | -0.6 |
| RDX | 0.004607 | 0.9 |
| Compressed Gas | 0.528353 | 100.0 |

The effects of increased (for oxy-acetylene or RDX) or decreased (for compressed gas) temperature on the volume of additional gases produced was not considered. The large differences between the number of moles of gas produced suggest that correcting for temperature effects would not change the main conclusion that the oxy-acetylene driven shock tube does not have a large jet effect.